\documentclass[conference]{IEEEtran}
\usepackage[pdftex]{graphicx}
\usepackage{amsmath}
\usepackage{nicefrac}
\usepackage{balance}
\usepackage{url}
\usepackage{pgfplots}
\usetikzlibrary{shapes.multipart,intersections}
\usepackage{cite}
\usepackage{amsmath,amssymb,amsfonts,steinmetz,bm}
\usepackage{mathtools,algpseudocode,algorithm,MnSymbol}
\usepackage{graphicx}
\usepackage{textcomp}
\usepackage{xcolor}
\def\BibTeX{{\rm B\kern-.05em{\sc i\kern-.025em b}\kern-.08em
    T\kern-.1667em\lower.7ex\hbox{E}\kern-.125emX}}
    
\usepackage{tikz}
\usetikzlibrary{arrows.meta}
\usetikzlibrary{calc}
\makeatletter
\newcommand{\gettikzxy}[3]{%
  \tikz@scan@one@point\pgfutil@firstofone#1\relax
  \edef#2{\the\pgf@x}%
  \edef#3{\the\pgf@y}%
}
\usepackage[nolist]{acronym}
\makeatother
\usepackage[draft]{todonotes}   
\usepackage{upgreek} 
\usepackage{seqsplit}
\usepackage{hyperref}
\usepackage{balance}
\usepackage[font={footnotesize}]{caption}
\usepackage{subcaption}
\usepackage[inline]{enumitem}
\newtheorem{rem}{Remark}

\acrodef{RIS}{reconfigurable intelligent surface}
\acrodef{BS}{base station}
\acrodef{UE}{user equipment}
\acrodef{LoS}{line-of-sight}
\acrodef{NLoS}{non-line-of-sight}
\acrodef{NF}{nearfield}
\acrodef{SNR}{signal-to-noise ratio}
\acrodef{SINR}{signal-to-interference-and-noise-ratio}
\acrodef{SISO}{single-input-single-output}
\acrodef{PEB}{position error bound}
\acrodef{FIM}{Fisher information matrix}
\acrodef{SDP}{semidefinite program}
\acrodef{PSD}{positive semidefinite}
\acrodef{LMI}{linear matrix inequality}
\acrodef{MC}{multi-carrier}
\acrodef{MIMO}{multiple inputs multiple outputs}
\acrodef{OEB}{orientation error bound}
\acrodef{DoD}{Direction of Departure}
\acrodef{TDoA}{Time Difference of Arrival}

\newcommand{\vect}[1]{\boldsymbol{#1}}

\begin{document}
\title{Constrained RIS Phase Profile Optimization and Time Sharing for Near-field Localization}
\author{\IEEEauthorblockN{Moustafa Rahal\IEEEauthorrefmark{1}\IEEEauthorrefmark{3}, Beno\^{i}t Denis\IEEEauthorrefmark{1}, Kamran Keykhosravi\IEEEauthorrefmark{2}, \\Musa Furkan Keskin\IEEEauthorrefmark{2}, Bernard Uguen\IEEEauthorrefmark{3}, Henk Wymeersch\IEEEauthorrefmark{2}}
\IEEEauthorblockA{\IEEEauthorrefmark{1}
CEA-Leti, Université Grenoble Alpes, F-38000 Grenoble, France\\
\IEEEauthorrefmark{2}
Department of Electrical Engineering, Chalmers University of Technology, Gothenburg, Sweden\\
\IEEEauthorrefmark{3}
Université Rennes 1, IETR - UMR 6164, F-35000 Rennes, France\\
}}
\maketitle

\begin{abstract} 
The rising concept of \ac{RIS} has promising potential for Beyond 5G localization applications. We herein investigate different phase profile designs at a reflective \ac{RIS}, which enable \acl{NLoS} positioning in \acl{NF} from downlink single antenna transmissions. 
We first derive the closed-form expressions of the corresponding \ac{FIM} and \ac{PEB}. Accordingly, we then propose a new localization-optimal phase profile design, 
assuming prior knowledge of the \acl{UE} location. Numerical simulations in a canonical scenario show that our proposal outperforms conventional \ac{RIS} random and directional beam codebook designs in terms of \ac{PEB}. We also illustrate the four beams allocated at the \ac{RIS} (i.e., one directional beam, along with its derivatives with respect to space dimensions) and show how their relative weights according to the optimal solution can be practically implemented through time sharing (i.e., considering feasible beams sequentially). 
\end{abstract}
\begin{IEEEkeywords}
\Acl{NF} localization, \acl{NLoS}, \ac{RIS} phase optimization.
\end{IEEEkeywords}
\acresetall
\section{Introduction}
Among the potential 6G technologies, \acp{RIS} stand out for their ability to purposely shape wireless environments~\cite{strinati2021}. 
A typical \ac{RIS} generally comprises a large number of controllable elements, 
which can be adjusted (typically, in terms of their phases) by means of lightweight electronics so as to behave as electromagnetic mirrors or lenses. 
\Acp{RIS} also have appealing applications to spatial awareness and sensing~\cite{WymeerschVehTechMag}, for instance to overcome \ac{LoS} blockages through newly induced multipath components, hence making localization feasible when conventional systems would fail. Beyond, they can be also beneficial to control and locally/timely improve localization accuracy when the \ac{LoS} path is present. Such flexibility and synergies between data communication and on-demand localization services are expected to be among the main drivers in future 6G systems, thus making \ac{RIS} a key enabling technology. 
Beyond quite extensive works on wireless localization exploiting the spherical wavefront of incoming signals at large receiving \acp{RIS} (e.g., \cite{hu_beyond_2018, alegria_cramer-rao_2019, guidi_radio_2020,abu2021near}), other recent research studies have also considered the use of passive reflective \acp{RIS} in the context of parametric multipath-aided positioning, 
in both \ac{LoS} (e.g., \cite{wymeersch_beyond_2020, Elzanaty2021, he_large_2020, zhang_towards_2020, keykhosravi2021siso}) and \ac{NLoS} (e.g., \cite{rahal2021ris, liu_reconfigurable_2020}) conditions. 

An important 
challenge with \ac{RIS} 
lies in the optimization of their profiles (i.e., their reflection coefficients and/or their phases), 
before or 
after performing channel estimation. For data communication purposes, the latter problem 
can be solved based on the estimated cascaded channel responses or using a priori location information~\cite{abrardo2020intelligent}. The former problem 
can be solved with a variety of approaches, including that maximizing recovery performance, 
based on the Discrete Fourier Transform (DFT) 
and Hadamard matrices,
or harnessing external location information too~\cite{hu2020location}.
%
 \begin{figure}
 \centering
 \includegraphics[width=0.8\linewidth]{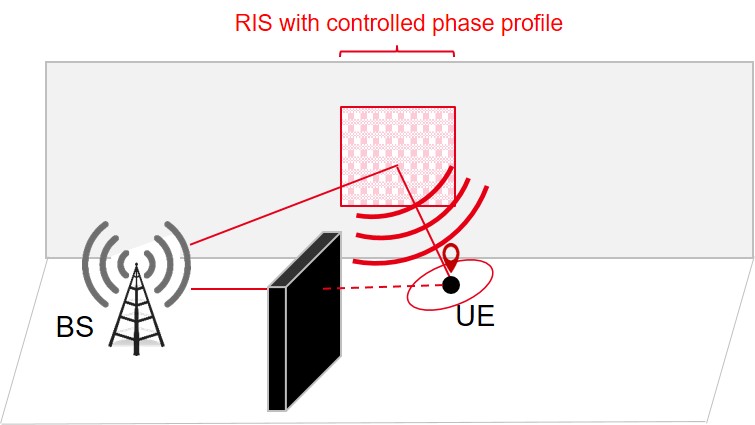}
 \caption{Typical \ac{NLoS} positioning scenario with one single reflective \ac{RIS} in \acl{NF} over \acl{SISO} downlink transmissions.}
 \label{fig:Scenario}
 \vspace{-6mm}
 \end{figure}
As for \acp{RIS}-based localization more specifically, in \cite{wymeersch_beyond_2020}, joint \ac{RIS} selection and directional reflection beam design has been considered while assuming prior knowledge of the \ac{UE} location, which intuitively corresponds to concentrating the reflected power towards the \ac{UE} and accordingly, increase \ac{SINR} at the \ac{UE}. 
Alternatively, the authors in \cite{rahal2021ris, keykhosravi2021siso} have used simpler random \ac{RIS} phase profiles for asynchronous positioning in a downlink \ac{SISO} \ac{MC} transmission context. This scheme does not require any prior information (neither about the channel, nor about the \ac{UE} location), but is not optimal under a priori \ac{UE} location information.
Finally, in \cite{Elzanaty2021}, RIS
phases and beamformers are jointly optimized with respect to both the \ac{PEB} and the \ac{OEB} in a generic \ac{MIMO} \ac{MC} context, based on a \ac{SNR} criterion. 


In this paper, in contrast to the previous contributions and leveraging the simpler \ac{SISO} downlink positioning scheme of \cite{keykhosravi2021siso,rahal2021ris}, we design suitable phase profiles at a reflective \ac{RIS} through \ac{PEB} optimization in an \ac{NLoS} context (See Fig.~\ref{fig:Scenario}),while putting more emphasis on practical implementation constraints (typically, forcing the RIS complex element response to lie on the unit-circle). Beyond making localization feasible (with significantly degraded accuracy in comparison with \ac{LoS} conditions though)~\cite{rahal2021ris}, the goal is to further optimize \ac{NLoS} positioning performance, while relying on direct localization using the \ac{RIS}-reflected path (i.e., as estimated at the \ac{UE} over downlink transmissions). 
Our main contributions are: \emph{(i)} we derive closed-form expressions of both the \acs{FIM} and the \ac{PEB}, while assuming a generic nearfield-compliant formulation for the \ac{RIS} response, \emph{(ii)} we show how the \ac{PEB} optimization problem can be solved efficiently and how localization-optimized \ac{RIS} phase profiles are obtained, considering practical time sharing among different profiles; 
\emph{(iii)} we compare the performance of the resulting localization-optimal phase profiles, along with its constrained (and thus sub-optimal) variants, with that of conventional random and directional designs, and \emph{(iv)} we discuss the shapes taken by the \ac{RIS} beam and its successive derivatives with respect to the 3D dimensions (in spherical coordinates) in light of our specific estimation problem.

\subsubsection*{Notations} {Vectors and matrices are denoted, respectively, with a lower-case and upper-case bold letter (e.g., $\vect{x}, \vect{X}$), while subscripts are used to denote their indices. Transpose, conjugate and hermitian conjugate are respectively denoted by $(.)^\top$, $(.)^{*}$ and $(.)^{\mathsf{H}}$. Furthermore, the operator $\mathrm{tr}(\vect{X})$ denotes the trace of matrix $\vect{X}$ and $\text{diag}(\vect{x})$ denotes a diagonal matrix with diagonal elements defined by vector $\vect{x}$.  Finally, $\Vert\cdot\Vert$ is the $l_{2}$-norm operator and $(.)^\smallstar$ denotes the solution of an optimization problem.
}

 
  \begin{figure}[t]
 \centering
 \includegraphics[width=0.75\linewidth]{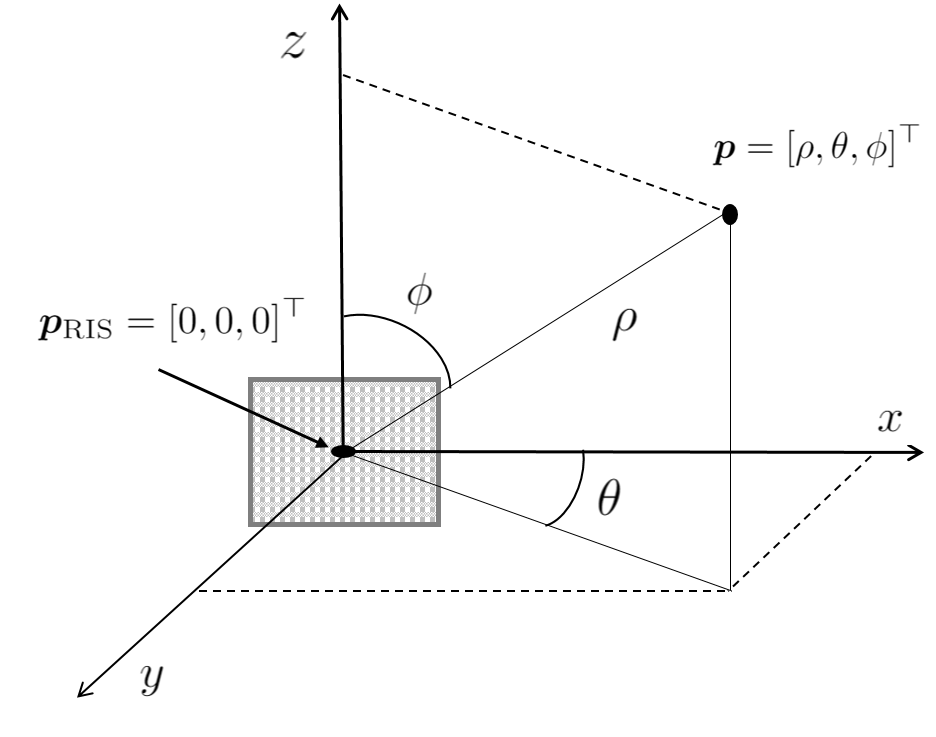}
 \caption{Problem geometry for a \ac{UE} in $\vect{p}$ and the \ac{RIS} phase center in $\vect{p}_{\text{RIS}}$, as the origin of both spherical and Cartesian coordinates systems. 
 }
 \vspace{-4mm}
 \label{fig:Geometry}
 \end{figure}

\section{System Model}
We consider a 3D localization setup consisting of a single-antenna \ac{BS}, a single-antenna \ac{UE} and a planar reflective RIS composed of $M$ elements. The corresponding 3D locations are expressed in the same global reference coordinates system as follows: $\vect{p}_{\text{\ac{BS}}} \in \mathbb{R}^{3 \times 1}$ is a vector containing the known \ac{BS} coordinates, \textcolor{black}{$\vect{p}_{\text{RIS}} \in \mathbb{R}^{3 \times 1}$} is a vector containing the known coordinates of the RIS center, \textcolor{black}{$\vect{p}_{m} \in \mathbb{R}^{3\times 1}$} is a vector containing the known coordinates of the $m$-th element \textcolor{black}{in the RIS}, and $\vect{p}\in \mathbb{R}^{3\times 1}$ is a vector containing \ac{UE}'s unknown coordinates, expressed as $[x,y,z]^{\top}$ in the Cartesian coordinates system or $[\rho, \theta, \phi]^{\top}$ in the spherical coordinates system, where $\rho$ is the range from the system's origin, $\theta$ is the azimuth angle measured from the positive $x$-axis and likewise, $\phi$ is the elevation angle from the positive $z$-axis. Without loss of generality, in the sequel, we choose the \ac{RIS} phase center as the origin of the coordinates system. The geometry of the problem is illustrated in Fig.~\ref{fig:Geometry}.
%
%
We consider a millimeter wave (mmWave) downlink communication scenario in presence of \ac{LoS} blockage, where the \ac{BS} broadcasts a narrowband pilot signal $x_{t} \in \mathbb{C}$ with a bandwidth $W$ over $T$ transmissions and transmit power $P_{\text{tx}}$. 
  In \ac{NLoS}, the complex signal $y_t \in \mathbb{C}$ received by the \ac{UE} at time $t$ after \ac{RIS} reflection is
\vspace{-1mm}
\begin{align}
    y_{t} & =\beta \vect{a}^{\top}(\vect{p})\vect{\Omega}_{t}\vect{a}(\vect{p}_{\text{BS}}) x_{t}  
    + n_{t}, \label{eq:Received_signal}
\end{align}
where $\vect{a}(.) \in \mathbb{C}^{M\times1}$ is a steering vector representing the RIS response,
  in its most generic formulation (i.e., encompassing the \ac{NF} regime like in~\cite{rahal2021ris}), whose $m$-th entry with respect to the $m$-th \ac{RIS} element $\vect{p}_{m}$ and the \ac{RIS} phase center $\vect{p}_{\text{RIS}}$ is 
\vspace{-1mm}
\begin{align}
    [\vect{a}(\vect{p})]_{m}=\exp\left(-\jmath\frac{2\pi}{\lambda}\left(\Vert\vect{p}-\vect{p}_{m}\Vert - \Vert\vect{p}-\vect{p}_{\text{RIS}}\Vert \right)\right).\label{eq:RIS_response}
\end{align}
Moreover, {the transmit symbol energy is defined as $E_{s} = \mathbb{E}\{|x_{t}|^{2}\} = \nicefrac{P_{\text{tx}}}{W} $ with total transmit energy $E_{\text{tot}} = E_{s}MT$,} {${n}_t \sim \mathcal{CN}({0},N_0)$ is the independent and identically distributed (i.i.d.) observation noise of power spectral density $N_0$}, $\beta$ is the time-invariant complex channel gain for the reflected path and $\vect{\Omega}_{t} = \text{diag}(\vect{\omega}_{t}) $ where $\vect{\omega}_{t} \in \mathbb{C}^{M\times 1}$ is the $t$-th phase profile vector applied across the $M$ \ac{RIS} elements. This received signal can hence be vectorized over $T$ transmissions into $\vect{y} \in \mathbb{C}^{T\times1}$ as follows\footnote{Without loss of generality, we assume a constant pilot $x_t=\sqrt{E_s}$ is transmitted.}
\vspace{-3mm}
\begin{align}
    \vect{y}=\sqrt{E_{s}}\beta \vect{F}^{\top} \vect{a}(\vect{p})   + \vect{n}, \label{eq:generalized_received_signal}
\end{align}
where $\vect{F}= [\vect{f}_1,\ldots,\vect{f}_T] \in \mathbb{C}^{M\times T}$ with $\vect{f}_t = \vect{\Omega}_t \vect{a}(\vect{p}_{\text{BS}}) \in \mathbb{C}^{M\times1}$, and $\vect{\Omega}_t$ taking its values in the set of valid\footnote{So-called valid profiles correspond to practically feasible complex values according to real \ac{RIS} hardware limitations (e.g., unit-modulus values with quantized phases).} RIS phase profiles:
\vspace{-3mm}
\begin{align}
    \left\vert [\vect{f}_t]_m \right\vert =1, \forall t,m. \label{eq:RISconstraint}
\end{align}

\section{Localization-Optimal RIS Profile Design}
In this section, we show how RIS profiles can be optimized to minimize the \ac{PEB} in a specific position. In Remark \ref{rem:perfectprior}, we address how the resulting chicken-and-egg problem can be resolved (as the goal of designing the RIS profiles is to localize the user, while the design itself requires knowledge of the user's position). 
\subsection{FIM and PEB}
We first define the vector of position and channel parameters in the 3D spherical coordinates system, as $\vect{\upzeta}_\text{sph}=[ \rho, \theta, \phi,  \beta_{r}, \beta_{i}]^\top \in \mathbb{R}^{5\times1}$, and compute the \acs{FIM} accordingly \cite[Chapter~3.7]{kay_fundamentals}
 \begin{align}
     \vect{J}_{\text{sph}}(\vect{\upzeta}_{\text{sph}}) =
    \frac{2E_s}{N_0}  \text{Re} \left\{
    \left(\frac{\partial \vect{\upmu}} {\partial\vect{\upzeta}_\text{sph}}\right)^{\mathsf{H}} \frac{\partial \vect{\upmu}} {\partial\vect{\upzeta}_{\text{sph}}}
   \right\} 
    \in \mathbb{R}^{5\times 5}, \label{eq:FIMsph}
 \end{align}
 where $\vect{\upmu} = \beta \vect{F}^{\top} \vect{a}(\vect{p})$  denotes the noiseless part of the observation. 
 To obtain the closed-form expressions of the \acs{FIM} terms, we differentiate $\vect{\upmu}$ with respect to the corresponding parameters
 \begin{align}
 \left[\frac{\partial \vect{\upmu}}{\partial {\rho}},\frac{\partial \vect{\upmu}}{\partial {\theta}},  \frac{\partial \vect{\upmu}}{\partial 
 {\phi}} \right]& =\beta \vect{F}^\top\left[ \Dot{\vect{a}}_{\rho}(\vect{p}), \Dot{\vect{a}}_{\theta}(\vect{p}), \Dot{\vect{a}}_{\phi}(\vect{p})\right]\\
     \left[\frac{\partial \vect{\upmu}}
    {\partial {{\beta}}_{r}},\frac{\partial \vect{\upmu}}
    {\partial {{\beta}}_{i}}\right] &= \vect{F}^\top \vect{a}(\vect{p}) [1,\jmath], 
\end{align}
 where $\Dot{\vect{a}}_{x}(\vect{p}) = \nicefrac{\partial \vect{a}(\vect{p})}
{\partial x}\in \mathbb{C}^{M\times 1}$.
Then, introducing $\vect{\upzeta}_\text{car}=[ \vect{p}^{\top}, \beta_{r}, \beta_{i}]^\top \in \mathbb{R}^{5\times1}$ as the set of position and channel parameters in Cartesian coordinates system, 
we use the Jacobian $\vect{C} = \nicefrac {\partial\vect{\upzeta}_{\text{sph}} } {\partial\vect{\upzeta}_{\text{car}}}$ to transform the previous \acs{FIM} into
\begin{align}\label{eq:carFIM}
    & \vect{J}_\text{car}(\vect{\upzeta}_\text{car}) = \vect{C}^{\top} \vect{J}_\text{sph}(\vect{\upzeta}_\text{sph}) \vect{C}.
\end{align}
Finally, we characterize the positioning performance by means of the \ac{PEB}, which is a lower bound on the accuracy of any unbiased location estimator, and is computed as \cite[Chapter~2.4.2]{vanTrees}
\vspace{-3mm}
\begin{align}
\mathrm{PEB}(\vect{F};\vect{\upzeta}_\text{car}) & = \sqrt{\mathrm{tr}\left(\left[\vect{J}_\text{car}^{-1}(\vect{\upzeta}_\text{car})\right] _{(1:3,1:3)} \right)}\label{eq:PEB}\\
& \le \sqrt{\mathbb{E}\left\{ \Vert\vect{p}-\hat{\vect{p}} \Vert^2\right\}},
\end{align}
where we have made the dependence of the precoding matrix $\vect{F}$ explicit.

\subsection{PEB Minimization}
Assuming prior knowledge of \ac{UE}'s position, we formulate the \ac{PEB} optimization problem under a total power constraint, as follows 
\vspace{-3mm}
\begin{subequations}
\begin{align} \label{eq:PEB_optimization}
     \min_{\vect{F}} \quad &  \mathrm{PEB}(\vect{F};\vect{\upzeta}_\text{car}) \\
     \text{s.t.} \quad & \mathrm{tr}(\vect{F}\vect{F}^{\mathsf{H}})=MT. 
\end{align}
\end{subequations}
We then suggest relaxing the above program: first, by using the change of variable $\vect{X}=\vect{FF}^{\mathsf{H}}$ and then, by removing the constraint $\text{rank}(\vect{X})=T$ \cite[Chapter~7.5.2]{boyd2004convex}\cite{garcia_optimal_2018}, 
yielding
\begin{subequations}
\begin{align}
     \min_{\vect{X,u}} \quad & \vect{1}^{\top}\vect{u} \\
     \text{s.t.} \quad 
     & 
     \begin{bmatrix}
    \vect{J}_\text{car} & \vect{e}_{k} \\ 
    \vect{e}_{k}^{\top} & u_k 
    \end{bmatrix} \succeq 0, k = 1, 2, 3 ,\\ 
     & \mathrm{tr}(\vect{X})=MT, \\ 
     & \vect{X} \succeq 0, 
\end{align}
\end{subequations}
where $\vect{u} = [u_{1}, u_{2}, u_{3}]^{\top}$ is an auxiliary variable and $\vect{e}_{k}$ is the $k$-th column of the identity matrix. This optimization problem is a convex \ac{SDP} since the \acs{FIM} is a linear function of $\vect{X}$, as we can see from \eqref{eq:FIMsph} to \eqref{eq:carFIM}, and the constraints are either \acp{LMI} or linear equalities. According to~\cite[Appendix~C]{4359542}, the optimal precoder covariance matrix $\vect{X}^{\smallstar}$ is 
of the form 
\begin{align}
    \vect{X}^{\smallstar} = \vect{U} \vect{\Lambda} \vect{U}^{\mathsf{H}} 
    \label{eq:Xopt}
\end{align}
where $\vect{\Lambda} \in \mathbb{C}^{4\times4}$ is a \ac{PSD} matrix, denoting by its diagonal entries the beam weights applied to (or equivalently, the relative powers allocated to) the columns of $\vect{U}$ while
\vspace{-1mm}
\begin{align}
    \vect{U} \triangleq [
\vect{a}^*(\vect{p}) \hspace{1mm}
\Dot{\vect{a}}^*_{\rho}(\vect{p}) \hspace{1mm}
\Dot{\vect{a}}^*_{\theta}(\vect{p}) \hspace{1mm}
\Dot{\vect{a}}^*_{\phi}(\vect{p})
], \label{eq:defU}
\end{align}
which are  \ac{RIS} steering vector and the successive derivative beams with respect to the spherical coordinates system components as that involved in \eqref{eq:FIMsph}. Note that the space spanned by the columns of $\vect{U}$ can also be spanned by 4 orthonormalized vectors, by applying the Gramm-Schmidt algorithm to the columns of $\vect{U}$, so that $\vect{U}^{\mathsf{H}} \vect{U}=M\vect{I}_4$. For the remainder of this paper, we will use these orthonormalized vectors, since the error bounds are function of the latter \cite{garcia_optimal_2018}. 
This allows us to write the constraint $\mathrm{tr}(\vect{X})=MT$ as $\mathrm{tr}(\vect{\Lambda})=T$. 
Applying the above transformation (i.e., from $\vect{X} \in \mathbb{C}^{M\times M}$ to $\vect{\Lambda} \in \mathbb{C}^{4\times 4}$), the computational complexity is then significantly reduced, and the new optimization problem can be simply stated as 
\vspace{-4mm}
\begin{subequations} \label{finalOpt}
\begin{align} 
    \min_{\vect{\Lambda,u}} \quad & \vect{1}^{\top}\vect{u} \\
     \text{s.t.} \quad 
     & 
     \begin{bmatrix}
    \vect{J}_\text{car} & \vect{e}_{k} \\ 
    \vect{e}_{k}^{\top} & u_k 
    \end{bmatrix} \succeq 0, k = 1, 2, 3 ,  \\
     & \mathrm{tr}(\vect{\Lambda})=T, \\
     & \vect{\Lambda} \succeq 0. 
\end{align}
\end{subequations}
Finally, the problem can be further relaxed by restricting $\vect{\Lambda}$ to be diagonal, i.e., $\vect{\Lambda}=\text{diag}(\vect{\lambda})$, in which case the entries in $\vect{\lambda}$ can be interpreted as power allocations or time units assigned to each column of $\vect{U}$. {To solve the optimization problem \eqref{finalOpt}, we used CVX \cite{cvx}}.

\begin{rem}[Assumption of prior knowledge] \label{rem:perfectprior}
In a real system, perfect 
a priori knowledge of the \ac{UE} location is not available, but can be reasonably approximated by the latest \ac{UE}'s estimated location (typically, while tracking the \ac{UE} in the steady-state regime). In other words, we make use of this prior information to optimize the operating conditions for the next UE location estimate, given that the UE would be quasi-static in the meantime. Note that the presumed location uncertainty associated to this prior (if only made available by an estimator, e.g., as an error covariance or an uncertainty ellipse) can be taken into account in our optimization problem (\ref{eq:PEB_optimization}), by  minimizing the worst-case PEB in a region around the estimated UE location (i.e., in a set of points rather than in a single point), like in \cite{keskin_optimal_2021}.  Furthermore, given a \ac{BS}--\ac{RIS} deployment, the optimization routine can be run offline and tabulated as a function of possible \ac{UE} locations, so that the \ac{RIS} profile can be reconfigured during the online phase at no extra computational cost, based on this location estimate.
\end{rem}

\subsection{Practical RIS Phase Profiles and Time Sharing}\label{subsec:Practical_Phase_Profiles}
When solving the optimization problem above, multiple approaches can be taken to generate RIS phase profiles that satisfy the constraint \eqref{eq:RISconstraint}. We limit our discussion to the case where $\vect{\Lambda}=\text{diag}(\vect{\lambda})$.  
\begin{itemize}
    \item \emph{Optimize, then constrain:} from the optimal value $\vect{X}^{\smallstar} = \vect{U}\vect{\Lambda}_{\text{opt}}\vect{U}^{\mathsf{H}}$ of \eqref{finalOpt}, 
    we transform  
    the orthonormal beams in $\vect{U}$ into their unit-modulus versions (using gradient projections as in \cite[Algorithm~1]{7849224})
    \item \emph{Constrain, then optimize:} in this approach, we first project the columns of $\vect{U}$ to satisfy the constraint \eqref{eq:RISconstraint} using the same method from  \cite[Algorithm~1]{7849224}, and then solve \eqref{finalOpt} with the corresponding set of non-orthonormal vectors. 
\end{itemize}
In either case, once \eqref{finalOpt} is solved and the set of precoders that satisfy \eqref{eq:RISconstraint} are determined, 
we aim to find time allocations $T_i$, $i\in \{1,2,3,4\}$ subject to $\mathrm{\sum}_{i=1}^{4}T_{i}=T$ and $T_i\in \mathbb{N}$, with $T_i \approx \lambda_i \in \mathbb{R}_{+}$. This problem can be solved by rounding $\lambda_i$ to the nearest integer. Moreover, a more general allocation for arbitrary $T$ can be found by solving 
\eqref{finalOpt} with $\mathrm{tr}(\vect{\Lambda})=1$, in which case, the values $\lambda_i\in [0,1]$ refer to the relative frequencies of the different RIS phase profiles. We can then set $T_i\approx \lambda_i T$, rounded to the nearest integer. With smaller $T$, the temporal quantization errors will become more pronounced, leading to PEB performance degradation. In particular, when $\lambda_i\ll 1$, the corresponding beam may never be selected, in which case the PEB will be infinite (since all 4 columns of $\vect{U}$ must be used). To address this,  we force each column to be used at least once as a RIS phase configuration. 


\section{Simulation Results}

\subsection{Simulation Parameters}
To assess the performance of the proposed design, several simulations have been performed in a canonical scenario, considering an indoor mmWave setting, using the parameters in Table \ref{table:params}. 
%
\begin{rem}[Practical estimators]
While our analysis is limited to performance bounds, it is expected that practical algorithms can achieve these bounds, when operating in a high \ac{SNR} regime. Low-complexity \ac{NF} localization methods were for instance discussed in \cite{abu2021near,rinchi2022compressive}. We also note that the proposed RIS phase profile designs involve only limited signaling or control overhead, since only the current user location needs to be provided to the RIS controller. 
\end{rem}

\begin{table}
\centering
\resizebox{\columnwidth}{!} {
\begin{tabular}{ |c|c||c|c| } 
 \hline
 parameter & value & parameter & value\\
 \hline
 $f_{c}$ & $28$ $\mathrm{GHz}$ & wavelength & $\approx 1.07$ $\mathrm{cm}$\\
 $W$ & $120$ $\mathrm{kHz}$ & UE loc. $\vect{p}$ & $[1, 1...15, 1]$ $\mathrm{m}$\\ 
 $N_0$ & $-174$ $\mathrm{dBm/Hz}$ &  BS loc. $\vect{p}_{\text{BS}}$ & $[5, 5, 0]$ $\mathrm{m}$\\ 
 noise figure $n_f$ & 8 $\mathrm{dB}$ &  RIS loc. $\vect{p}_{\text{RIS}}$ & $[0, 0, 0]$ $\mathrm{m}$\\ 
 $P_{\text{tx}}$ & 20 $\mathrm{dBm}$  & RIS size & $M = 32\times 32$ elements\\
  $E_{\text{tot}} $& $(\nicefrac{P_{\text{tx}}}{W})MT$  & transmissions & $T = 40$ \\
 \hline
\end{tabular}\vspace{-4mm}
}
\caption{General simulation parameters.}
    \label{table:params}
    \vspace{-4mm}
\end{table}
\subsection{Performance Analysis}
\subsubsection{Visualization of Beams}
\begin{figure}
     \centering
     \begin{subfigure}[b]{1\columnwidth}
         \centering
         \resizebox{1\columnwidth}{!}{\input{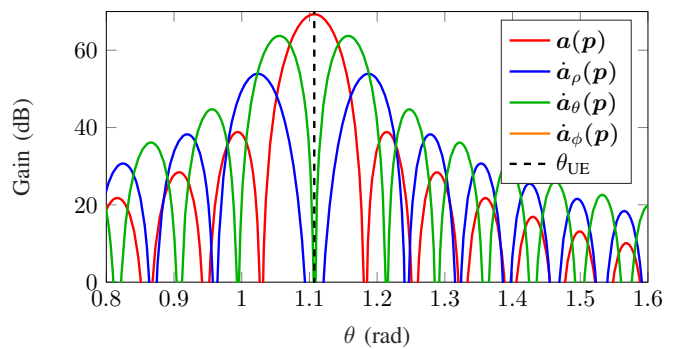}}
         \caption{Beam responses vs. $\theta$ with $\theta_{\text{UE}}$ denoting the UE azimuth angle.}
         \label{fig:OrthoNormBeams-}
     \end{subfigure}
     \hfill
     \vspace{0.5mm}
     \begin{subfigure}[b]{1\columnwidth}
         \centering
         \resizebox{1\columnwidth}{!}{\input{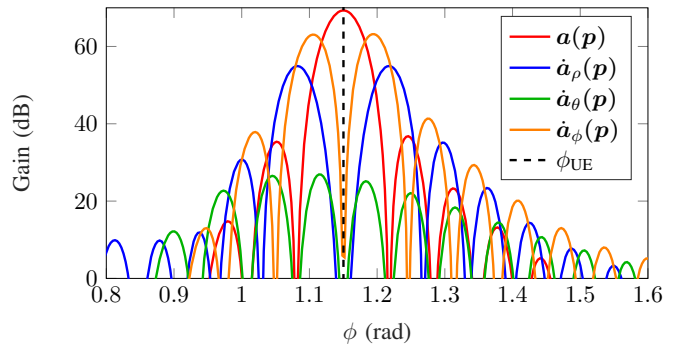}}
         \caption{Beam responses vs. $\phi$ with $\phi_{\text{UE}}$ denoting the UE elevation angle.}
         \label{fig:OrthoNormBeams-b}
     \end{subfigure}
     \hfill
     \vspace{0.5mm}
     \begin{subfigure}[b]{1\columnwidth}
         \centering
          \resizebox{1\columnwidth}{!}{\input{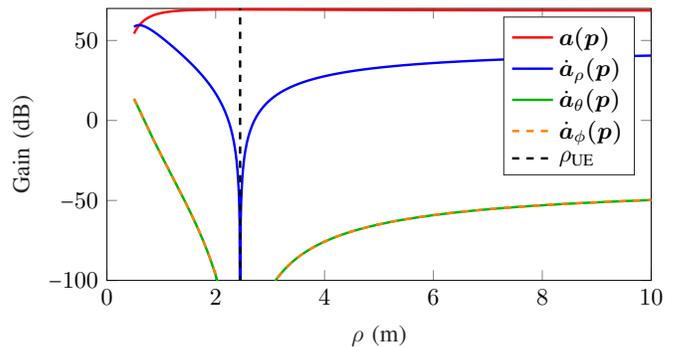}}
         \caption{Beam responses vs. $\rho$ with $\rho_{\text{UE}}$ denoting the UE range to the RIS.}
         \label{fig:OrthoNormBeams-c}
     \end{subfigure}
    \caption{Orthonormalized beams applied at the reflective RIS as a function of the three polar coordinates (incl. the directional beam and its derivatives).} 
    \label{fig:OrthoNormBeams} \vspace{-4mm}
\end{figure}

In Fig.~\ref{fig:OrthoNormBeams} (a)--(c), we first show the four orthonormalized beams applied at the reflective RIS (i.e., the directional beam and its derivatives), as a function of the 3 spherical coordinates.
{To elaborate, the figure displays a so-called \emph{Gain}, which corresponds to the expression $|\vect{u}^\top\vect{a}( \tilde{\vect{p}}) |^{2}$, where $\vect{u}$ represents a column in $\vect{U}$ and $\tilde{\vect{p}}$  is an arbitrary test location. The columns of $\vect{U}$ are generated according to \eqref{eq:defU} with $\vect{p} = [1,2,1]^\top$. 
%
For visualization purposes, we show cuts of possible positions $\tilde{\vect{p}}$ along the three polar coordinates, fixing the other two coordinates to the corresponding values of $\vect{p} = [1,2,1]^\top$.
}
As expected and similar to \cite{keskin_optimal_2021}, it is  noticed that the beam derivatives get their null values at the actual UE range/direction and that each beam null is visible in the three dimensions. These nulls lead to significant variation around the position to be estimated, thereby improving the positioning accuracy. In particular, the variation with respect to $x \in \{\theta,\phi,\rho\}$ is most pronounced for the beam $\dot{\vect{a}}_x(\vect{p})$.


\subsubsection{Comparison of Design Strategies}
\begin{figure}
 \centering
 \resizebox{1\columnwidth}{!}{
%
%
\definecolor{mycolor1}{rgb}{1.00000,0.00000,1.00000}%
\definecolor{mycolor2}{rgb}{0.00000,1.00000,1.00000}%
\begin{tikzpicture}

\begin{axis}[%
width=10cm,
height=7.4cm,
at={(0.758in,0.481in)},
scale only axis,
xmin=1,
xmax=15,
xtick = {0, 1, 2, 3, 4, 5, 6, 7, 8, 9, 10, 11, 12, 13, 14, 15},
xlabel style={font=\color{white!15!black}},
xlabel={Distance (m)},
ymode=log,
ymin=0.01,
ymax=10,
yminorticks=true,
ylabel style={font=\color{white!15!black}},
ylabel={PEB (m)},
axis background/.style={fill=white},
  legend columns=1, 
legend style={at={(0.35,0.4)}, anchor=north west, legend cell align=left, align=left, draw=white!15!black}
]
\addplot [color=orange, dashed, line width = 1.2pt]
  table[row sep=crcr]{%
1.73 	0.0377988868237765\\
2.44 	0.129914823613078\\
3.31 	0.284688681321228\\
4.24 	0.51408246116443\\
5.19 	0.962794436053798\\
6.16 	1.43396373848958\\
7.14 	2.55722298755813\\
8.12 	3.41872256902243\\
9.11 	5.15519065332483\\
10.09	7.43919785495137\\
11.09	9.6046206144409\\
12.08	12.0628261162577\\
13.07	14.590025597303\\
14.07	17.4986051426528\\
15.06	23.7855690910926\\
};
\addlegendentry{Random phase, $T=40$}

\addplot [color=cyan, dashdotted , line width = 1.2pt ]
  table[row sep=crcr]{%
1.73	0.0264149785399135\\
2.44	0.0738853537892368\\
3.31	0.193511961275596\\
5.24	0.680689793338596\\
6.16	0.994857297727342\\
7.14	1.84791018718749\\
8.12	2.58256895304234\\
9.11	3.81386480905224\\
10.09	5.13774500806478\\
11.09	6.46103213605145\\
12.08	7.97670746426841\\
13.07	10.6815364137261\\
14.07	11.9834059293311\\
15.06	16.203631025964\\
};
\addlegendentry{Random phase, $T=80$}

\addplot [color=red, dashed, line width = 1.2pt]
  table[row sep=crcr]{%
1.73 	0.0235843086447941\\
2.44 	0.0690019147385876\\
3.31 	0.172986960669428\\
4.24 	0.342012073953034\\
5.19 	0.609151363389478\\
6.16 	0.975072233933024\\
7.14 	1.70113457022664\\
8.12 	2.40487282290592\\
9.11 	3.41216267575695\\
10.09	4.35226678156324\\
11.09	5.30302609560589\\
12.08	7.05590054482577\\
13.07	10.512626013101\\
14.07	12.1780292870965\\
15.06	14.1954526010019\\
};
\addlegendentry{Random phase, $T=100$}

\addplot [color=black!30!green, line width = 1.2pt, mark=o, mark options={solid, black!30!green}]
  table[row sep=crcr]{%
1.73 	0.00374064530607768\\
2.44 	0.0119740682021535\\
3.31 	0.0419964628668895\\
4.24 	0.130840469094646\\
5.19 	0.347091899505488\\
6.16 	0.801529474784911\\
7.14 	1.65805461403451\\
8.12 	3.14453542696482\\
9.11 	5.56361324194025\\
10.09	9.30349927464035\\
11.09	14.8487692270903\\
12.08	22.7911577041817\\
13.07	33.8403528221001\\
14.07	48.8347910049683\\
15.06	68.7524519527575\\
};
\addlegendentry{Directional beams, $T=40$, $r=0.5$ m}

\addplot [color=black, line width = 1.2pt, mark=o, mark options={solid, black}]
  table[row sep=crcr]{%
1.73 	0.031432522084465\\
2.44 	0.0449465630745092\\
3.31 	0.130152488411264\\
4.24 	0.236491942360989\\
5.19 	0.354020270146411\\
6.16 	0.470009927429934\\
7.14 	0.581538831457874\\
8.12 	0.691964425944073\\
9.11 	0.837316326623363\\
10.09	1.00137979084378\\
11.09	1.23201931992545\\
12.08	1.57691480696431\\
13.07	2.01266823271177\\
14.07	2.45356162880533\\
15.06	2.85550189678455\\
};
\addlegendentry{Directional beams, $T=40$, $r=2$ m}

\addplot [color=blue, line width = 1.2pt]
  table[row sep=crcr]{%
1.73205080756888	0.00129804094427493\\
2.44948974278318	0.00350919437329026\\
3.3166247903554	0.00821758899848704\\
4.24264068711929	0.0166255316616386\\
5.19615242270663	0.0299507052149414\\
6.16441400296898	0.0494112612994798\\
7.14142842854285	0.0762263843804851\\
8.12403840463596	0.111615969323717\\
9.1104335791443	0.156800308596591\\
10.0995049383621	0.212999948170776\\
11.0905365064094	0.281435562625183\\
12.0830459735946	0.363327928907574\\
13.076696830622	0.459897869561563\\
14.0712472794703	0.572366255333785\\
15.0665191733194	0.701953990776547\\
};
\addlegendentry{Optimized, no constraint, $\vect{\Lambda}$ full}

\end{axis}

\end{tikzpicture}%
}
 \caption{PEB comparison as a function of the RIS-UE distance, for optimized RIS phase profiles (with optimal $\vect{\Lambda}$), random RIS phase profiles and directional RIS beams (for different numbers of transmissions $T$ and uncertainty levels $r$).} 
 \label{fig:AggregatedResponseOptimalLambda}
 \vspace{-6mm}
 \end{figure}
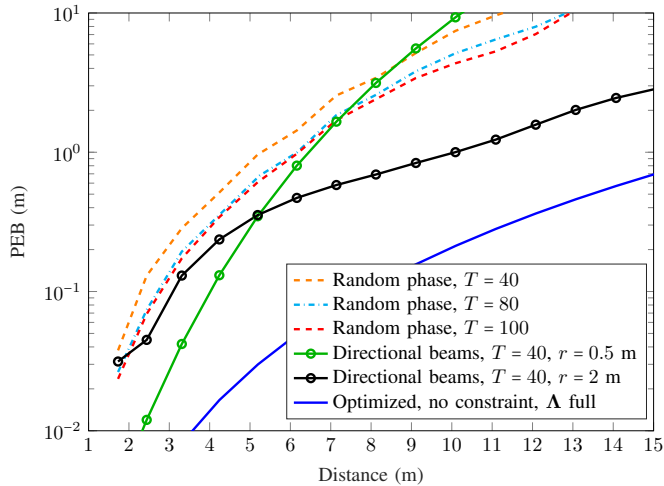

In Fig.~\ref{fig:AggregatedResponseOptimalLambda}, as a function of the RIS-UE distance, we then compare the PEB achievable with the optimal design (i.e., under unconstrained $\vect{\Lambda}$) with that obtained with purely random RIS phase profiles  \cite{wymeersch_beyond_2020} and directional RIS beams \cite{abu2021near}. In the latter approach, directional RIS beams are generated uniformly distributed into a sphere centered around the actual UE position, while assuming different levels of uncertainty (i.e., different values for the sphere radius $r$). 
The total energy was fixed to make the comparison fair among the different schemes. 
We first observe that the optimal design systemically outperforms the two other designs, whatever the distance. As for random phase profiles more specifically, beyond a certain number of transmissions (say, around 80 in the shown example), no further spatial diversity can be brought into the problem by the new profiles (i.e., most of the space has already been covered by previous profiles) so that performance asymptotically reaches a limit. With directional beams, the effect of prior UE location uncertainty is rather remarkable. Smaller uncertainty (i.e., $r=0.5$ m) indeed provides much better results at short distances (even close to the optimal design) but conversely leads to poor angular diversity and hence becomes counterproductive at long distances, just as if a unique beam was always selected to point in the same direction over all the transmissions. 

\subsubsection{Feasible RIS Profiles}
\begin{figure}
 \centering
 \resizebox{1\columnwidth}{!}{
%

\begin{tikzpicture}

\begin{axis}[%
width=10cm,
height=7.4cm,
at={(0.758in,0.481in)},
scale only axis,
xmin=1,
xmax=15,
xtick = {0, 1, 2, 3, 4, 5, 6, 7, 8, 9, 10, 11, 12, 13, 14, 15},
xlabel style={font=\color{white!15!black}},
xlabel={Distance (m)},
ymode=log,
ymin=0.01,
ymax=1,
yminorticks=true,
ylabel style={font=\color{white!15!black}},
ylabel style={font=\color{white!15!black}},
ylabel={PEB (m)},
axis background/.style={fill=white},
legend style={at={(0.35,0.4)}, anchor=north west, legend cell align=left, align=left, draw=white!15!black}
]
\addplot [color=black, line width = 1.2pt, dashed,  mark=o , mark options={solid, black}]
  table[row sep=crcr]{%
1.73205080756888	0.00213855175702521\\
2.44948974278318	0.00485361775144284\\
3.3166247903554	0.0104688046581099\\
4.24264068711929	0.0208722452805204\\
5.19615242270663	0.0373924070727976\\
6.16441400296898	0.0620980692209128\\
7.14142842854285	0.0950460977457891\\
8.12403840463596	0.139056274994808\\
9.1104335791443	0.195713658043837\\
10.0995049383621	0.266002442840578\\
11.0905365064094	0.351583690317003\\
12.0830459735946	0.454595806336389\\
13.076696830622	0.57655710626718\\
14.0712472794703	0.718411522527471\\
15.0665191733194	0.881309016043934\\
};
\addlegendentry{Constrain, then optimize, $\vect{\Lambda}$ diagonal}

\addplot [color=orange,  line width = 1.2pt]
  table[row sep=crcr]{%
1.73205080756888	0.00355994152243129\\
2.44948974278318	0.00871116975329999\\
3.3166247903554	0.019978765545159\\
4.24264068711929	0.0400941487920261\\
5.19615242270663	0.0723309109378145\\
6.16441400296898	0.119279624257761\\
7.14142842854285	0.184560188279044\\
8.12403840463596	0.270601017297028\\
9.1104335791443	0.380496813069938\\
10.0995049383621	0.517202941256162\\
11.0905365064094	0.683170424970784\\
12.0830459735946	0.881219085851935\\
13.076696830622	1.11451711653733\\
14.0712472794703	1.38656923017819\\
15.0665191733194	1.70069365226746\\
};
\addlegendentry{Time division ($T=4$)}

\addplot [color=blue,  line width = 1.2pt]
  table[row sep=crcr]{%
1.73205080756888	0.00231655220426963\\
2.44948974278318	0.00555945449016278\\
3.3166247903554	0.0126570576686988\\
4.24264068711929	0.0253946410260281\\
5.19615242270663	0.0457992446343258\\
6.16441400296898	0.0755584626020134\\
7.14142842854285	0.116832450664564\\
8.12403840463596	0.171290213460042\\
9.1104335791443	0.240890047453976\\
10.0995049383621	0.327445436125739\\
11.0905365064094	0.432550121208989\\
12.0830459735946	0.558093871282094\\
13.076696830622	0.706103515845309\\
14.0712472794703	0.878713278033947\\
15.0665191733194	1.07789865677781\\
};
\addlegendentry{Time division ($T=8$)}

\addplot [color=green,  line width = 1.1pt]
  table[row sep=crcr]{%
1.73205080756888	0.00225030304778553\\
2.44948974278318	0.00496222862665053\\
3.3166247903554	0.0111382677914408\\
4.24264068711929	0.0223366057528422\\
5.19615242270663	0.0402593630765446\\
6.16441400296898	0.0664744612017143\\
7.14142842854285	0.102648763299547\\
8.12403840463596	0.15047619242828\\
9.1104335791443	0.211680842255558\\
10.0995049383621	0.28776073120371\\
11.0905365064094	0.380190132990558\\
12.0830459735946	0.490796734950684\\
13.076696830622	0.621416959372929\\
14.0712472794703	0.773773353906336\\
15.0665191733194	0.949371837861508\\
};
\addlegendentry{Time division ($T=16$)}

\addplot [color=red,  line width = 1.2pt]
  table[row sep=crcr]{%
1.73205080756888	0.00217325288010667\\
2.44948974278318	0.00496222862665053\\
3.3166247903554	0.0107623487085858\\
4.24264068711929	0.0215557643334071\\
5.19615242270663	0.0388250702772573\\
6.16441400296898	0.0641389896412406\\
7.14142842854285	0.0989773020677006\\
8.12403840463596	0.145071096715416\\
9.1104335791443	0.204071915698883\\
10.0995049383621	0.277411070988564\\
11.0905365064094	0.366518766494902\\
12.0830459735946	0.473174215133851\\
13.076696830622	0.599146385329917\\
14.0712472794703	0.746069823258122\\
15.0665191733194	0.915382268613572\\
};
\addlegendentry{Time division ($T=32$)}


\end{axis}

\end{tikzpicture}%
}
 \caption{Achievable PEB as a function of the RIS-UE distance with both optimal diagonal $\vect{\Lambda}$  and its practical implementation through time sharing depending on $T$ (i.e., so-called ``Time division", allocating $T_i$ over $T$ transmissions for the $i$-th feasible beam separately, with related $\vect{f}_i$ forced onto the unit circle).}
 \label{fig:TimeSharingPEBVsTime}
 \vspace{-4mm}
 \end{figure}
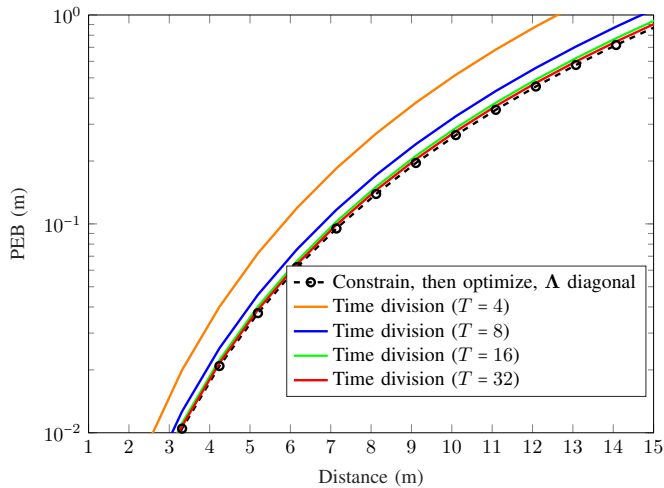

Fig.~\ref{fig:PEBwrtSolvers} provides a benchmark of the achievable PEB as a function of the distance with different variants of the problem solver, as described in Section ~\ref{subsec:Practical_Phase_Profiles}, assuming both full and diagonal $\vect{\Lambda}$, considering both \emph{constrain, then optimize}, and \emph{optimize, then constrain} approaches. 
Forcing $\vect{\Lambda}$ to be diagonal in the optimization  has almost no effect on the results in comparison with unconstrained $\vect{\Lambda}$. This is likely due to the fact that the initial beam vectors in $\vect{U}$ are orthogonal and hence, $\vect{\Lambda}$ shall be structurally quasi-diagonal accordingly.
Constraining before optimization turns out to be superior over constraining after optimization. Hence, we will consider the former approach from now on.
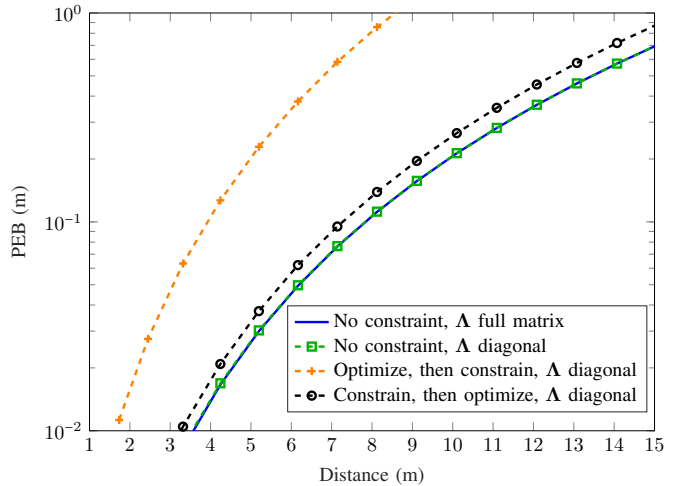
\begin{figure}
 \centering
 \resizebox{1\columnwidth}{!}{
%
\begin{tikzpicture}

\begin{axis}[%
width=10cm,
height=7.4cm,
at={(0.758in,0.481in)},
scale only axis,
xmin=1,
xmax=15,
xtick = {0, 1, 2, 3, 4, 5, 6, 7, 8, 9, 10, 11, 12, 13, 14, 15},
xlabel style={font=\color{white!15!black}},
xlabel={Distance (m)},
ymode=log,
ymin=0.01,
ymax=1,
yminorticks=true,
ylabel style={font=\color{white!15!black}},
ylabel={PEB (m)},
axis background/.style={fill=white},
legend style={at={(0.35,0.3)}, anchor=north west, legend cell align=left, align=left, draw=white!15!black}
]
\addplot [color=blue, line width = 1.2pt]
  table[row sep=crcr]{%
1.73205080756888	0.00129804094427493\\
2.44948974278318	0.00350919437329026\\
3.3166247903554	0.00821758899848704\\
4.24264068711929	0.0166255316616386\\
5.19615242270663	0.0299507052149414\\
6.16441400296898	0.0494112612994798\\
7.14142842854285	0.0762263843804851\\
8.12403840463596	0.111615969323717\\
9.1104335791443	0.156800308596591\\
10.0995049383621	0.212999948170776\\
11.0905365064094	0.281435562625183\\
12.0830459735946	0.363327928907574\\
13.076696830622	0.459897869561563\\
14.0712472794703	0.572366255333785\\
15.0665191733194	0.701953990776547\\
};
\addlegendentry{No constraint, $\vect{\Lambda}$ full matrix}


\addplot [color=black!30!green, dashed, line width = 1.2pt, mark=square, mark options={solid, black!30!green}]
  table[row sep=crcr]{%
1.73205080756888	0.00134665676264279\\
2.44948974278318	0.00365779002946062\\
3.3166247903554	0.00842713585314941\\
4.24264068711929	0.0168762798725076\\
5.19615242270663	0.0302318706078926\\
6.16441400296898	0.0497155990828684\\
7.14142842854285	0.0765486488668858\\
8.12403840463596	0.111952259280723\\
9.1104335791443	0.157147695411216\\
10.0995049383621	0.213356205359702\\
11.0905365064094	0.281798986120971\\
12.0830459735946	0.363697215744385\\
13.076696830622	0.46027197074409\\
14.0712472794703	0.572744370295222\\
15.0665191733194	0.70233549100067\\
};
\addlegendentry{No constraint, $\vect{\Lambda}$ diagonal}

\addplot [color=orange, dashed, line width = 1.2pt, mark=+, mark options={solid, orange}]
  table[row sep=crcr]{%
1.73205080756888	0.0112575235478894\\
2.44948974278318	0.0275471375047947\\
3.3166247903554	0.0631784039611863\\
4.24264068711929	0.126788831028493\\
5.19615242270663	0.228730423798264\\
6.16441400296898	0.37719529110347\\
7.14142842854285	0.583630560351329\\
8.12403840463596	0.855715551817268\\
9.1104335791443	1.20323657173563\\
10.0995049383621	1.63553930690796\\
11.0905365064094	2.16037457297271\\
12.0830459735946	2.78665942890331\\
13.076696830622	3.52441257950253\\
14.0712472794703	4.38471690086853\\
15.0665191733194	5.37806554335571\\
};
\addlegendentry{Optimize, then constrain, $\vect{\Lambda}$ diagonal}

\addplot [color=black, line width = 1.2pt, dashed, mark=o, mark options={solid, black}]
  table[row sep=crcr]{%
1.73205080756888	0.00213855175702521\\
2.44948974278318	0.00485361775144284\\
3.3166247903554	0.0104688046581099\\
4.24264068711929	0.0208722452805204\\
5.19615242270663	0.0373924070727976\\
6.16441400296898	0.0620980692209128\\
7.14142842854285	0.0950460977457891\\
8.12403840463596	0.139056274994808\\
9.1104335791443	0.195713658043837\\
10.0995049383621	0.266002442840578\\
11.0905365064094	0.351583690317003\\
12.0830459735946	0.454595806336389\\
13.076696830622	0.57655710626718\\
14.0712472794703	0.718411522527471\\
15.0665191733194	0.881309016043934\\
};
\addlegendentry{Constrain, then optimize, $\vect{\Lambda}$ diagonal}

\end{axis}

\end{tikzpicture}%
}
 \caption{Achievable PEB as a function of the distance for different variants of the problem solver.}
 \label{fig:PEBwrtSolvers}
 \vspace{-3mm}
 \end{figure}
 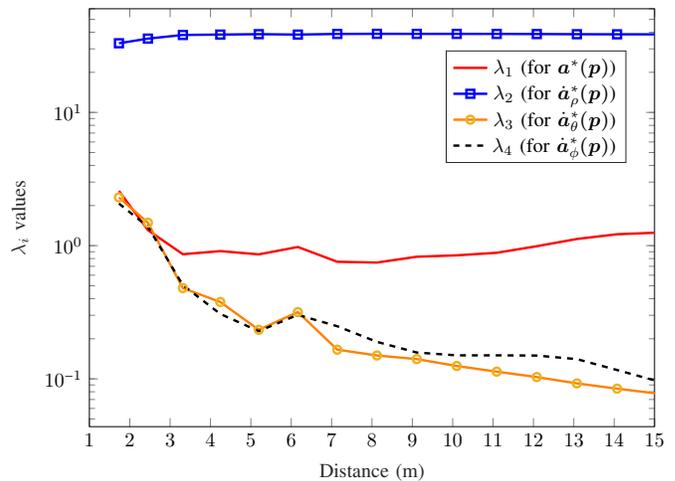
\begin{figure}
 \centering
 \resizebox{1\columnwidth}{!}{
%
%
\definecolor{mycolor1}{rgb}{0.00000,0.44700,0.74100}%
\definecolor{mycolor2}{rgb}{0.85000,0.32500,0.09800}%
\definecolor{mycolor3}{rgb}{0.92900,0.69400,0.12500}%
\definecolor{mycolor4}{rgb}{0.49400,0.18400,0.55600}%
\begin{tikzpicture}

\begin{axis}[%
width=10cm,
height=7.4cm,
at={(0.758in,0.481in)},
scale only axis,
xmin=1,
xmax=15,
xtick = {0, 1, 2, 3, 4, 5, 6, 7, 8, 9, 10, 11, 12, 13, 14, 15},
xlabel style={font=\color{white!15!black}},
xlabel={Distance (m)},
ymin=0,
ymax=60,
ymode=log,
ylabel style={font=\color{white!15!black}},
ylabel={$\lambda_i$ values},
axis background/.style={fill=white},
legend style={at={(0.95,0.9)}, anchor=north east, legend cell align=right, align=right, draw=white!15!black}
]
\addplot [color=red, line width = 1.2pt]
  table[row sep=crcr]{%
1.73205080756888	2.57535527920906\\
2.44948974278318	1.30905920760458\\
3.3166247903554	0.861925965108586\\
4.24264068711928	0.909783692500965\\
5.19615242270663	0.860039274973837\\
6.16441400296898	0.976880881934675\\
7.14142842854285	0.756840750754928\\
8.12403840463596	0.746326359059905\\
9.1104335791443	0.82561070855368\\
10.0995049383621	0.845729564608568\\
11.0905365064094	0.883830027775191\\
12.0830459735946	0.990225277550238\\
13.076696830622	1.12195722721841\\
14.0712472794703	1.21769808395272\\
15.0665191733194	1.25185797072694\\
};
\addlegendentry{$\lambda_1$ (for $\vect{a}^*(\vect{p})$)}

\addplot [color=blue, line width = 1.2pt, mark=square]
  table[row sep=crcr]{%
1.73205080756888	33.0530210922385\\
2.44948974278318	35.8169161004326\\
3.3166247903554	38.1532503667153\\
4.24264068711928	38.4053693070934\\
5.19615242270663	38.6784896796426\\
6.16441400296898	38.4031956112496\\
7.14142842854285	38.8292064726664\\
8.12403840463596	38.9143082379368\\
9.1104335791443	38.8759922384385\\
10.0995049383621	38.8792227769815\\
11.0905365064094	38.852719963779\\
12.0830459735946	38.7570503231098\\
13.076696830622	38.6449169779823\\
14.0712472794703	38.5813673027108\\
15.0665191733194	38.574369308178\\
};
\addlegendentry{$\lambda_2$ (for $\Dot{\vect{a}}^*_{\rho}(\vect{p})$)}

\addplot [color=orange, line width = 1.2pt, mark=o, mark options={solid, mycolor3}]
  table[row sep=crcr]{%
1.73205080756888	2.30194033060807\\
2.44948974278318	1.48669793821667\\
3.3166247903554	0.479855037868523\\
4.24264068711928	0.37769296427161\\
5.19615242270663	0.233437885117716\\
6.16441400296898	0.31733043861766\\
7.14142842854285	0.165650956268811\\
8.12403840463596	0.149779397356245\\
9.1104335791443	0.14073711459407\\
10.0995049383621	0.124925452253904\\
11.0905365064094	0.113280414369226\\
12.0830459735946	0.103121078637266\\
13.076696830622	0.0924652236610832\\
14.0712472794703	0.0844207718602874\\
15.0665191733194	0.0775794268697436\\
};
\addlegendentry{$\lambda_3$ (for $\Dot{\vect{a}}^*_{\theta}(\vect{p})$)}

\addplot [color=black, dashed, line width = 1.2pt]
  table[row sep=crcr]{%
1.73205080756888	2.06968329792815\\
2.44948974278318	1.38732675156922\\
3.3166247903554	0.504968599618363\\
4.24264068711928	0.307154025279182\\
5.19615242270663	0.228033009350862\\
6.16441400296898	0.302593055817186\\
7.14142842854285	0.248301799025154\\
8.12403840463596	0.189585996979044\\
9.1104335791443	0.157659928939342\\
10.0995049383621	0.15012220288214\\
11.0905365064094	0.150169592598699\\
12.0830459735946	0.149603320478948\\
13.076696830622	0.14066057096125\\
14.0712472794703	0.116513841024103\\
15.0665191733194	0.0961932941371217\\
};
\addlegendentry{$\lambda_4$ (for $\Dot{\vect{a}}^*_{\phi}(\vect{p})$)}

\end{axis}

\end{tikzpicture}%
}
 \caption{Diagonal terms of $\vect{\Lambda}_{\text{opt}}$ accounting for the different beam weights, as a function of the RIS-UE distance.}
 \label{fig:OptimizerLambdaOutputs}
 \vspace{-4mm}
 \end{figure}
Fig.~\ref{fig:OptimizerLambdaOutputs} shows the corresponding output of the optimizer in terms of $\lambda_{i}$ values, as a function of the RIS--UE distance. We notice that there is a strong dependence on the derivative beam with respect to the range. Moreover, as the UE moves away from the \ac{RIS}, we see less reliance on the angular derivative beams whose weights become nearly negligible, but almost a uniform dependence on directional.
Finally. in Fig.~\ref{fig:TimeSharingPEBVsTime}, we show the PEB achievable with our optimized design as a function of the RIS-UE distance, for both optimal $\vect{\Lambda}$  and its practical implementation through time sharing (depicted as ``Time division" here, see Section \ref{subsec:Practical_Phase_Profiles}) for distinct $T$ values, where the $i$-th beam is such that $\vect{f}_i$ is forced to lie onto the unit circle and used over $T_i$ transmissions out of $T$. As expected, lower values of $T$ lead to worse performance, as the temporal quantization effect is more pronounced. Inversely, as the value of $T$ increases the performance approaches, asymptotically, that of the optimized $\vect{\Lambda}$.

\subsection{Complexity}

Despite the possibility to generate and tabulate offline PEB-optimal RIS profiles as a function of the UE location so as to reduce the online computational cost (See Remark \ref{rem:perfectprior}), we now assess how the optimization problem scales with RIS size $M$. Calculating both directional and derivative beams has a complexity of $\mathcal{O}(M)$. Projecting the beams to the unit-modulus space has also a complexity of $\mathcal{O}(M)$. More precisely, this step has a complexity $\mathcal{O}(NM/\epsilon)$, where $N$ is the number of 3D points chosen in ~\cite[Algorithm~1]{7849224}, and $\epsilon$ is the accuracy limit for the algorithm convergence. Calculating the optimal $\vect{\Lambda}$ in our case is independent of $M$. So overall, the complexity is of order $\mathcal{O}(M)$, which is the same as that of directional codebooks. Beyond, for practical beamforming anyway, one does not need to calculate explicitly $\vect{X}$ but just to repeat beams based on optimized terms in $\vect{\Lambda}$.
%
%
\section{Conclusions}
In this paper, we have described a reflective \ac{RIS} phase profile design minimizing the PEB of \ac{NLoS} localization over downlink \ac{SISO} narrowband transmissions, while considering a generic near-field formalism for the \ac{RIS} response. On this occasion, we have shown that the theoretical optimal solution would involve the combination of four weighted beams at the \ac{RIS}, whose practical performance in terms of achievable PEB has been evaluated while considering more realistic unit-modulus beams. Finally, for the sake of implementability, we have also introduced a  time sharing scheme, assuming the application of each feasible beam sequentially, with very limited performance degradation whenever the overall number of transmissions is sufficiently large.
Future work will consider the evaluation of practical point estimators and tracking algorithms that use the proposed RIS phase profiles, the approximation of the required sequential beams under real reflective RIS hardware characterization \cite{rahal2022beam}, as well as the extension to multi-user and multi-RIS contexts. 

\section*{Acknowledgment}
This work has been supported, in part, by the EU H2020 RISE-6G project under grant 101017011 and by the MSCA-IF grant 888913 (OTFS-RADCOM).

\balance
\bibliographystyle{ieeetr}
\bibliography{references}

\end{document}